\documentclass[prb,twocolumn,superscriptaddress]{revtex4}
\usepackage{graphicx}
\usepackage{dcolumn}
\usepackage{bm}

\begin{document}
\title{On the role of backauditing for tax evasion in an agent-based
Econophysics model}
\author{G. Seibold}
\affiliation{Institute of Physics, BTU Cottbus, PBox 101344,
         03013 Cottbus, Germany}
\author{M. Pickhardt}
\affiliation{Institute of Economics, BTU Cottbus, PBox 101344,
         03013 Cottbus, Germany}

\begin{abstract}
We investigate an inhomogeneous Ising model in the context of
tax evasion dynamics where different types of agents are parametrized 
via local temperatures and magnetic fields. In particular, we analyse 
the impact of backauditing and endogenously determined penalty rates 
on tax compliance. Both features contribute to a microfoundation of 
agent-based econophysics models of tax evasion.   

\end{abstract}

\maketitle

\section{Introduction}
One of the first approaches to theoretically account for tax compliance
was given by Allingham and Sandmo \cite{alling72}, which incorporates
tax rates, potential penalties and audit probabilities as basic 
parameters in order to evaluate the behavior of expected utility maximizing tax payers.
However, it was realized early on that one of the major shortcomings
of the Allingham and Sandmo theory is the prediction of much too low
levels of tax compliance, as actually observed in industrialized nations.
It is believed that these shortcomings are partially due to the 
insufficient consideration of interaction dynamics among the actors of the 
tax compliance game, i.e., tax payers, tax advisors, tax authorities, and tax law makers. 
Another reason might be that many analyses have not incorporated 
backauditing. Although Allingham and Sandmo already considered backauditing 
(or lapse of time effects) as a possible extension of their basic theory, the issue of
backauditing has been largely neglected in the literature. Noteable exceptions are 
Alm {\it et al.} \cite{alm93}, Antunes {\it et al.} \cite{antu07} 
and Hokamp and Pickhardt \cite{hopi10}.

It is for these reasons that agent-based models have been
set up as a comparatively new tool for analyzing tax compliance issues. 
In fact, an essential feature of any agent-based model 
is the direct non-market based interaction among agents, which is combined with 
some process that allows for changes in individual behavior patterns. 
Therefore, agent-based tax evasion models may be categorized according to 
the features of this individual interaction process. In econophysics 
models this process is commonly described within the Ising model
\cite{ising} where examples include Zaklan et al. 
\cite{zak08,zak09}, Lima and Zaklan \cite{lz08}, and Lima \cite{lima10}.
 
In contrast, if the interaction process is driven by parameter changes that 
induce behavioral changes via a utility function and (or) 
by stochastic processes that do 
not have physical roots, these models belong to the economics domain. 
Examples include Mittone and Patelli \cite{mp00}, Davis et al. \cite{dav03}, 
Bloomquist \cite{blom04,blom06}, Korobow et al. \cite{koro07}, 
Antunes et al. \cite{antu07}, Szab\'o et al. \cite{szabo09}, Meder et al. \cite{meder10}, 
Hokamp and Pickhardt \cite{hopi10}, of which 
some are summarized by Bloomquist \cite{blom06}
and Pickhardt and Seibold \cite{pick12}. 

In all agent-based tax evasion models  
the actual patterns and levels of tax evasion depend on 
two additional factors: the network structure of society and the tax 
enforcement mechanism. The network structure is implemented by alternative 
lattice types and tax enforcement consists of the two economic standard 
parameters audit probability and penalty rate.

In the present paper we study the effect of alternative backauditing
schemes on tax evasion within an multi-agent econophysics model.
Moreover, backauditing also enables us to incorporate endogenously determined penalty rates. 
Both features, backauditing and endogenous penalties, contribute to the 
microfoundation of 
econophysics models and, therefore, allow for a more realistic 
modelling of tax compliance behavior within this modelling frame.   

We present our model and formalism in Sec. \ref{sec2}. Results are analyzed
in Sec. \ref{sec3}, where we discuss first the case of homogeneous societies
in Sec. \ref{secresa}, which is then compared with the tax evasion 
of multi agent societies in Sec. \ref{secresb}. We conclude our considerations
in Sec. \ref{sec4}.  

\section{Model and Formalism}\label{sec2}
Our considerations are based on the Ising model, described by the following
hamiltonian,
\begin{equation}\label{eq:isi}
H=\sum_{ij}J_{ij}S_i S_j + \sum_i B_i
\end{equation}
where $J_{ij}$ describes the coupling of Ising variables (spins) $S_i=-1$,$1$ 
between lattice sites $R_i$ and $R_j$. 
In the present context $S_i=1$ is interpreted as a compliant
tax payer and $S_i=-1$ as a non-compliant one. 
We implement the model 
on a two-dimensional $1000 \times 1000$ square lattice with 
nearest-neighbor interactions
$J_{ij}\equiv J$ for $|R_i-R_j|=1$ (lattice constant $a\equiv 1$). 
We expect that our results are robust with regard to variations of 
the network structure in analogy to the investigations of Zaklan {\it et al.}
Ref. \cite{zak08}.
Eq. (\ref{eq:isi}) contains also
the coupling of the spins to a local magnetic field $B_i$
which together with a local temperature $T_i$ distinguishes the
behaviorally different types of agents. Concerning the latter we assume that
it is imposed by the coupling of each lattice site $i$ to a heat-bath 
with temperature $T_i$ \cite{note1}, as if it were part of a canonical
ensemble.  We then use the heat-bath 
algorithm [cf. \cite{krauth}] in order to evaluate statistical 
averages of the model. 
The probability for a spin at lattice site $i$ to take the values 
$S_i=\pm 1$ is given by 
\begin{equation}\label{eq:prob} 
p_i(S_i)=\frac{1}{1+\exp\{-[E(-S_i)-E(S_i)]/T_i\}} 
\end{equation}
and $E(-S_i)-E(S_i)$ is the energy change for a spin-flip at site $i$.
Upon picking a random number $0 \le r \le 1$ the spin takes the value
$S_i=1$ when $r < p_i(S_i=1)$ and $S_i=-1$ otherwise.
One time step then corresponds to a complete sweep through the lattice. 
We note that a generalization of the model with regard to the incorporation of
non-equilibrium dynamics has been recently proposed in Refs. \cite{lima10,lima121,lima122}.

Following Ref. \cite{hopi10} we consider societies which are composed 
of the following four types of agents:
(i) \textit{selfish a-type agents}, which take advantage from 
non-compliance and, thus, are characterized by $B_i/T_i < 0$ and 
$|B_i| > J$; (ii) \textit{copying b-type agents}, which copy tax behavior 
of their social environment or neighborhood. This can be modelled
by $B_i << J$ and $J/T_i \gtrsim 1$; 
(iii) \textit{ethical c-type agents}, which are
practically always compliant and which are parametrized by 
$B_i/T_i > 0$ and $|B_i| > J$; 
(iv) \textit{random d-type agents} which are in principle like c-types, 
but due to some confusion caused by tax law complexity act by chance within a 
certain range.
We implement this behavior by $B_i<<J$ and $J/T_i <<1$. 
Here and in the following all parameters are measured with respect 
to $J \equiv 1$. 
Note that with regard to the previous definitions the analysis in 
Ref. \cite{zak09} corresponds to homogeneous societies
of b- and d-type agents. Moreover, the model in Ref. \cite{lz08}
studies homogeneous societies of a-type and c-type agents.
Our aim instead is the investigation of heterogeneous societies
and the influence of different backaudit schemes and endogenously 
determined penalty rates on the extend of tax evasion.

We consider first the case where the detection of an evading agent 
in the current period enforces its compliance over the 
following $h$ time steps or periods. Note that in econophysics models of tax evasion 
$h$ is regarded as the penalty rate. The aforementioned procedure has been
invoked in Refs. \cite{zak08, zak09, lima10, pick12} 
and also implemented in a randomized variant in Ref. \cite{lz08}.
In Ref. \cite{lz08} $h$ has been interpreted as 
the time over which an agent is ashamed and feels guilty about his
behavior after detection. An alternative interpretation would be definite audits
by the tax authorities over $h$ time steps after an agent has been detected 
for the first time, a scheme that is known as conditional future 
auditing (Alm {\it et al.} \cite{alm93}).
In addition, we study the situation where an 
audit may also allow for screening the agent's tax declaration over 
several years in the past (backaudit). The basic idea is that
a backaudit is only performed by the tax authorities upon
reasonable initial suspicion, which in our model corresponds to 
tax evasion in the current period. In the literature this scheme is
known as conditional backauditing (see Alm {\it et al.} \cite{alm93}).
We therefore introduce a probability $p_a$
with which an audit is performed at a given lattice site (agent). 
If tax evasion is detected in the current period, 
the backaudit comprises also an
inspection of the preceding $b_p$ time steps. Denote with $n_e$ the
number of time steps over which the agent was evading during the
backaudited periods plus the current period. Then the number of future periods $k$, over which the 
agent is forced to be compliant, is set to $k=n_e*h$. 
Thus, the penalty rate $k$ is now endogenously determined.
Note, however, that the above limit of a fixed number of enforced future compliance 
periods $h$ is recovered in the limit of no backaudit $b_p=0$.

\section{Results}\label{sec3}
In this section we will first present results for the case of homogeneous
societies consisting of either a, b, c- or d-type agents. 
This allows for extracting
their specific behavior patterns under the three different audit schemes 
which we consider: 
(i) $b_p=0$ with penalty rate $h=4$, (ii) $b_p=5$ with penalty rate $k$, (iii) $b_p=10$ with penalty rate $k$.
It turns out that agent behavior is partially modified in a heterogeneous society 
due to interaction dynamics between different agent groups.

\subsection{Homogeneous societies}\label{secresa}
\begin{figure}[h!]
\begin{center}
\includegraphics[width=8cm,clip=true]{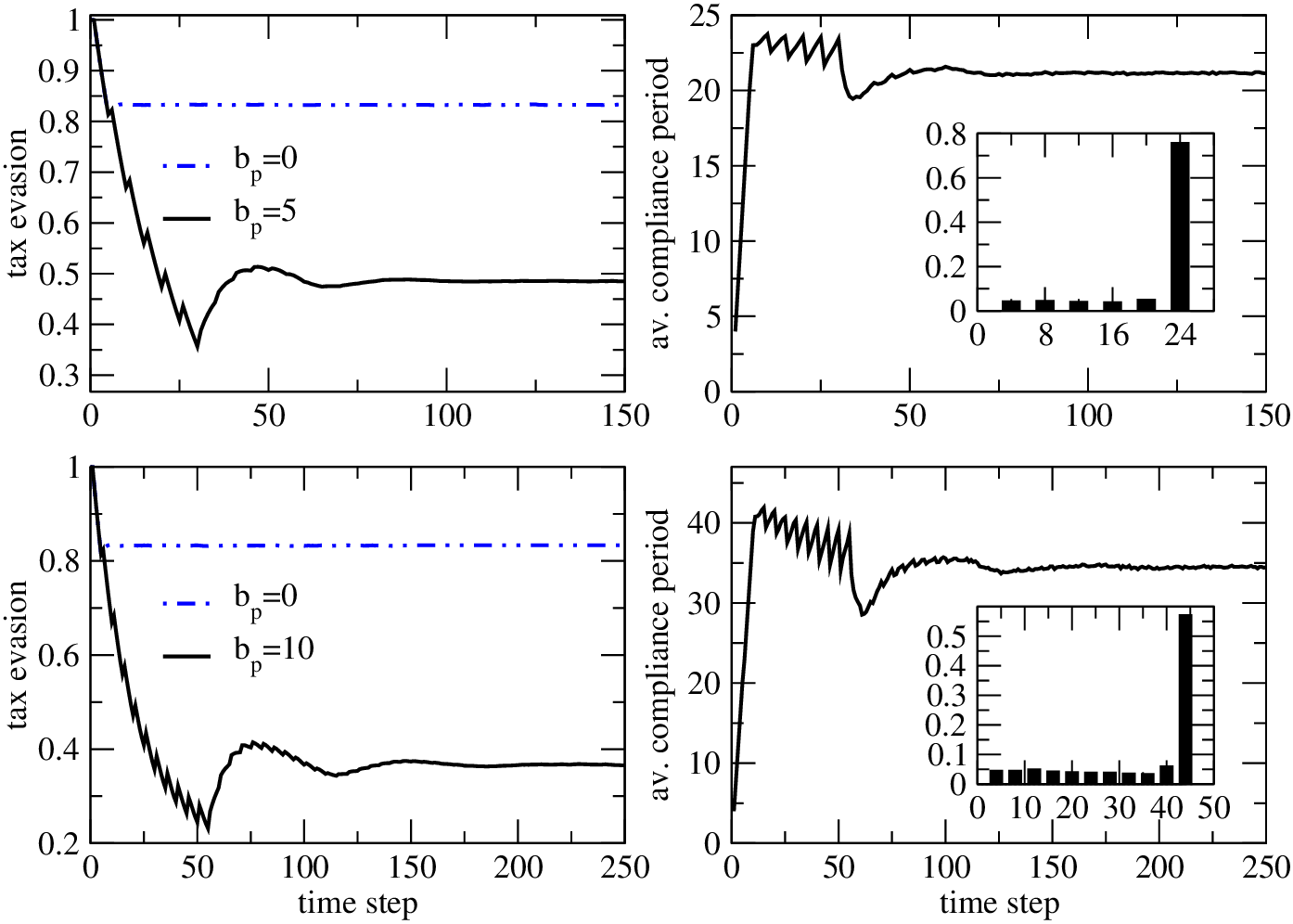}
\end{center}
\caption{Tax evasion dynamics for a society of $a$-type agents (temperature
$T=5$, external field $B=-15$). The upper (lower) panels display tax evasion for
 backaudit period $b_p=5$ ($b_p=10$). In both panels the dashed line corresponds 
to $b_p=0$ and $h=4$. In all cases the audit probability is set to
$p_a=5\%$.}
\label{fig1}
\end{figure}

Fig. \ref{fig1} displays the tax evasion dynamics for  a homogeneous 
system of selfish $a$-type agents. The endogenous preference with regard to 
non-compliance is incorporated by a negative field $B=-15$, which
is large in magnitude both with regard to the exchange coupling $J\equiv 1$
and the temperature scale $T=5$.

As initial condition we set the value of all agents to $-1$, i.e. $100\%$
tax evasion \cite{note2}. 
The upper left panel of Fig. \ref{fig1} compares the time
evolution for the extend of tax evasion in case of no backauditing ($b_p=0$) 
and a 
fixed compliance period $h=4$ (audit probability $p_a=5\%$) with backauditing 
($b_p=5$). Note that in the latter case the maximum penalty
is $k=6*4=24$ since the current period is included in the counting
 of total evasion periods $n_e$.
Up to time step $t_i=6$ the dynamics of both methods are undistinguishable,
whereas for $t_i>6$ tax evasion without backauditing ($b_p=0$)  rapidly
saturates in contrast to the curve with backauditing.
In fact, since we start from $100\%$ tax evasion it is unlikely that
for the small $p_a$ we are considering an agent which has been forced 
to compliance, is audited. 
Therefore, in both methods the extend of tax evasion is initially reduced 
in each time step by $p_a$, i.e. decreasing to $95\%$ , $90\%$ , $85\%$ ,
$80\%$ in time step $t_i=2$, $3$, $4$ , $5$.
Agents detected in period $t_i=1$ can
start to cheat again in period $t_i=6$. In case of no backauditing 
this fraction of evading agents will be almost compensated by
the newly detected evadors which drives the system into the stationary
state. On the other hand, backauditing ($b_p=5$) forces agents
which have been detected in $t_i > 1$ to stay compliant over a larger
period so that the corresponding curve continues to decrease. 
The following small upward spikes occur when these agents turn to non-compliance again.
The decrease continues until agents can acquire the maximum compliance 
period of $24$ time steps. This is the case after time step $t_i=30$ where 
agents which have been convicted in period $t_i=6$ return to non-compliance.
As a consequence the tax evasion probability starts to increase again
and slowly approaches a stationary state.
The upper right main panel illustrates the time evolution of the
average compliance period of detected agents. As anticipated, it rapidly
increases within the first $6$ time steps and converges to a stationary
value of $k \approx 21$ after $\sim 80$ time steps. Note that the average
$k$ is smaller than the maximum value of $k=24$ due to detected agents
which have been compliant within previous $b_p=5$ time steps. The corresponding
distribution is shown in the inset to the upper right panel of  Fig. \ref{fig1}.
Clearly, most of the detected agents ($\sim 75\%$)  are penalized with a 
compliance period of $k=24$, whereas the the remaining $25\%$ are forced
to compliance over $4, 8, 12, 16, 20$ time steps with almost equal
probability of $\sim 5\%$.
The lower two panels of Fig. \ref{fig1} analyze the analogous situation
for the larger backaudit period of $b_p=10$. Clearly this reduces even more
the percentage of tax evasion with respect to no backauditing,
but on the other hand increases the transient oscillation towards the
stationary state. 

\begin{figure}[h!]
\begin{center}
\includegraphics[width=8cm,clip=true]{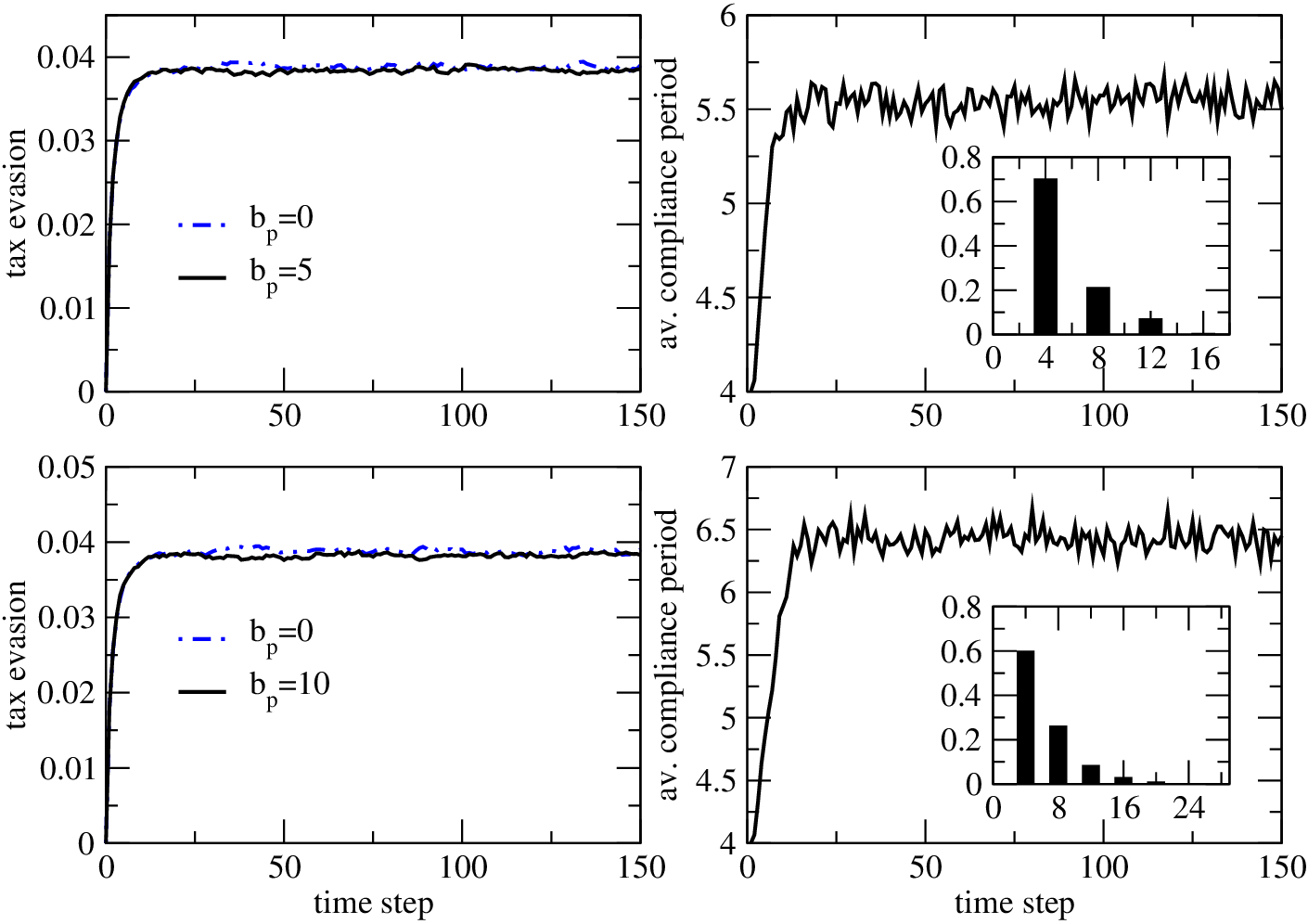}
\end{center}
\caption{Same as Fig. \ref{fig1} but for a society of $b$-type agents 
(temperature $T=2$, external field $B=0$).}
\label{fig2}
\end{figure}

Fig. \ref{fig2} displays the situation for a homogeneous society of
copying b-type agents which are characterized by $B=0$ and $T=2$, i.e. the
temperature is below the ordering transition of the two-dimensional
Ising model $T_c \approx 2.269$. Therefore, b-type agents are
guided by a line of action which copies the behavior of its neighbors
(i.e. sites which are coupled by $J \equiv 1$). 
As initial condition all agents are set to full compliance (tax evasion $0 \%$).
Due to the small but finite temperature some of the agents occasionally
change to non-compliance.
However, the small audit probability implies also a small probability
that  agents are detected twice or more often within $b_p$ time steps.
Therefore, for both backaudit periods $b_p=5,10$ the majority of agents
is sconced with a compliance
period of $k=4$ (cf. insets to the right panels of Fig. \ref{fig2})
so that the average compliance period is only slightly larger than
in case of no backaudit $b_p=0$ (cf. main right panels).
As a result, one obtains a small stationary value of tax evasion of $\sim 4\%$
after $\sim 15$ time steps where the curves with and without backaudit
are practically indistinguishable (cf. left panels
of Fig. \ref{fig2}).
It should be noted that the stationary value is {\it independent}
of the initial condition. Even in case of $100 \%$ tax evasion
in time step $t_i=1$ the detected agents have a small
probability to return to non-compliance resulting in the same
stationary value after a somewhat longer transient oscillation
of $\sim 30-40 $ time steps.

\begin{figure}[h!]
\begin{center}
\includegraphics[width=8cm,clip=true]{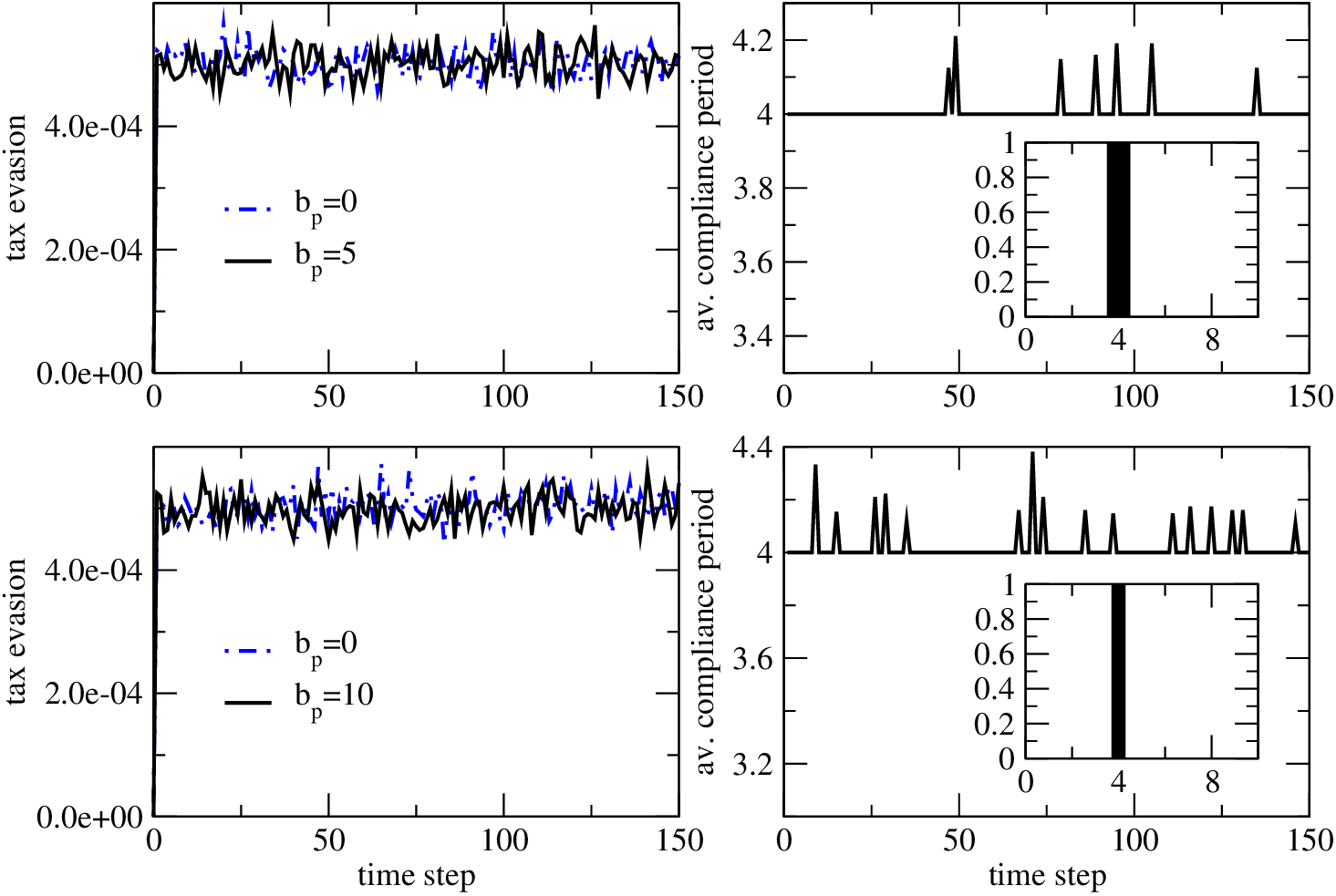}
\end{center}
\caption{Same as Fig. \ref{fig1} but for a society of $c$-type agents 
(temperature
$T=5$, external field $B=+15$).}
\label{fig3}
\end{figure}

The case of ethical c-type agents is shown in Fig. \ref{fig3}. These
are modeled by a strong positive field $B=15$ (i.e. $B/J > 1$ and
$B/T > 1$ with $T=5$) which enforces compliant behavior.
This strong field suppresses the probability for a behavioral
change towards non-compliance. From the right panels of Fig. \ref{fig3}
one can see that the average compliance period is dominantly $k=4$
(cf. insets),
with incidental spikes when some of the agents are detected twice within
$b_p$ time steps. Therefore, the number of spikes increases upon
increasing the backaudit period from $b_p=5$ to $b_p=10$.
In any case, backauditing has only a minor effect on c-type agents and 
the tax evasion rapidly
approaches a rather small stationary value of $\sim 5\cdot 10^{-4}$,
which would tend to zero in the limit $B/J, B/T \to \infty$.

\begin{figure}[h!]
\begin{center}
\includegraphics[width=8cm,clip=true]{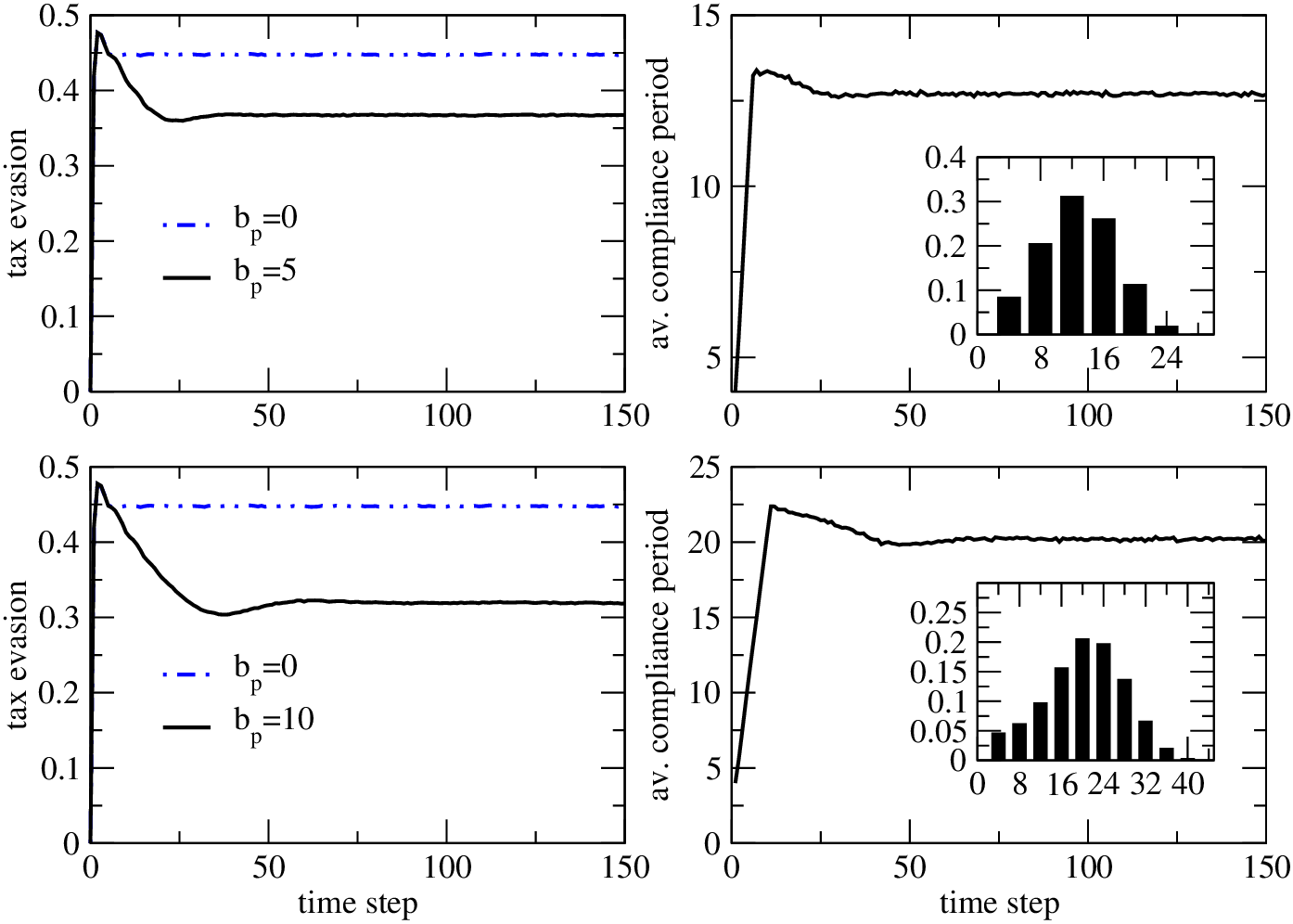}
\end{center}
\caption{Same as Fig. \ref{fig1} but for a society of $d$-type agents (temperature
$T=20$, external field $B=0$).}
\label{fig4}
\end{figure}

Finally, Fig. \ref{fig4} displays the case of a homogeneous society of
d-type agents which act independently from other agents and, thus, are 
characterized by a large temperature $T=20$ and $B=0$.
Setting as initial condition all agents to compliance, the large
temperature induces a change to non-compliance for almost half of the
society within the first two time steps (cf. left panels of Fig. \ref{fig4}).
Then the audit mechanisms drive the system again towards the stationary
states, which in case of d-type agents is different for the cases
with and without backauditing. In fact, since d-type agents act independently
we expect them on average to be half of the time compliant and
non-compliant, respectively. The backaudit within $b_p+1$ time steps
(i.e. including the current period) thus leads to an
average enforced compliance period of $k=h\cdot(b_p+1)/2$. This agrees
with the stationary values of the average compliance curves shown in the 
right panels of Fig. \ref{fig4} and the actual distribution is shown
in the insets. The increased compliance period under backauditing 
results in a smaller stationary
value of tax evasion, as can be seen in the left panels of Fig. \ref{fig4}.
Naturally, this value decreases with the length of the backaudit period
$b_p$.
    
\subsection{Heterogeneous societies}\label{secresb}
We continue our investigations by considering societies which contain
all four behaviorly different types of agents. In order to account for 
heterogeneity also within the individual agent groups, 
we allow for parameter variations concerning the 
characterizing temperature $T_{min} \le T \le T_{max}$ and fields
$B_{min} \le B \le B_{max}$.
Table \ref{tab1} specifies the parameter windows used in the following analysis
which correspond to a flat distribution around the same mean values 
used for the homogeneous societies in the previous subsection.

\begin{figure}[h!]
\begin{center}
\includegraphics[width=8cm,clip=true]{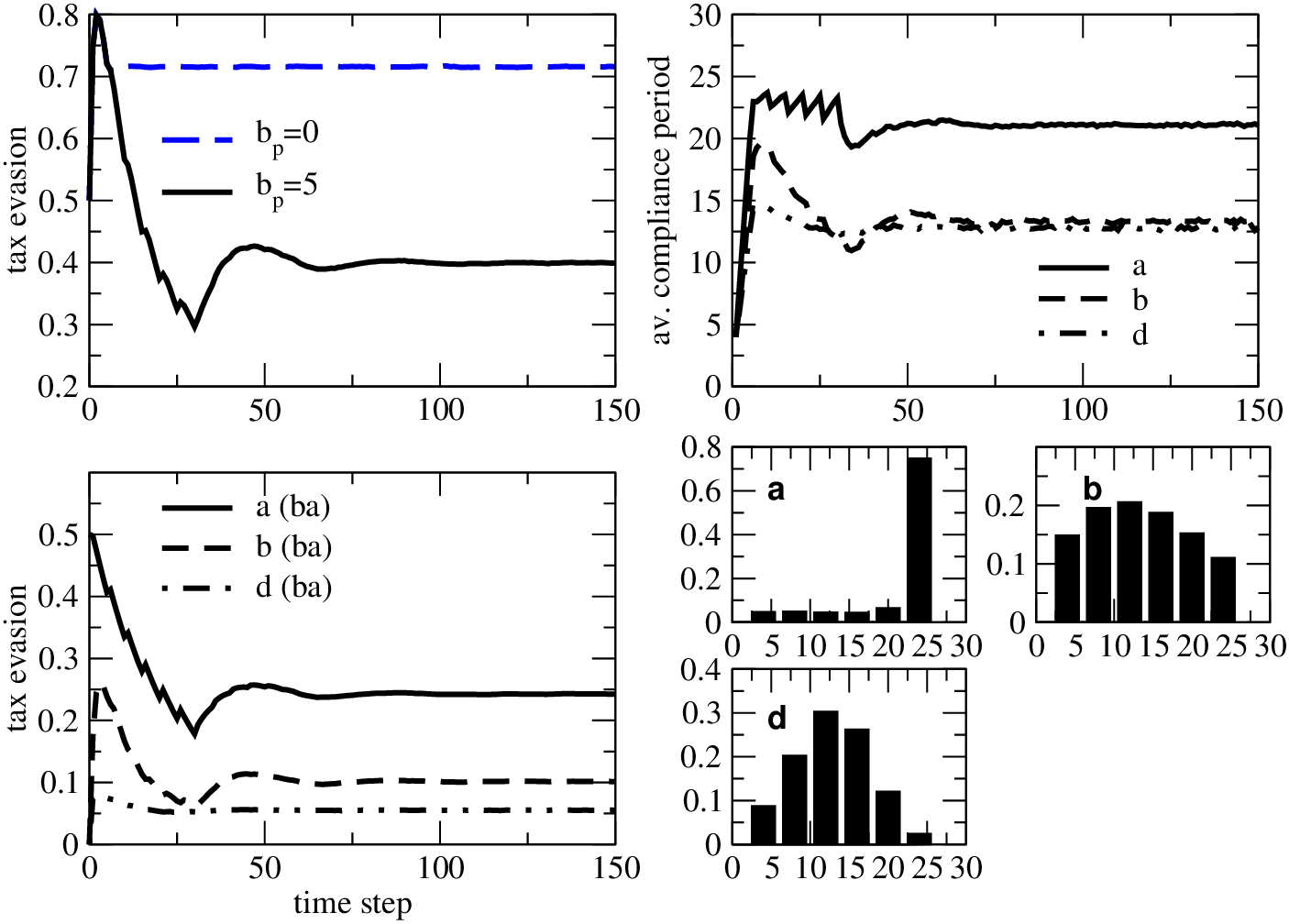}
\end{center}
\caption{Tax evasion dynamics for a society consisting of $50\%$ a-type, 
$35\%$ b-type, and $15\%$ d-type agents. The upper left panel compares
the extend of tax evasion and the dynamics of tax evasion for backaudit periods $b_p=0$ and $b_p=5$.
The lower left panel breaks down the evasion probability to the individual
agent types in case of backaudit period $b_p=5$. The upper right panel displays the average forced
compliance period (or penalty) for the individual agent types and the lower right panels
show the corresponding distribution. Audit probability is $p_a=5\%$ in each case.}
\label{fig5}
\end{figure}

\begin{figure}[h!]
\begin{center}
\includegraphics[width=8cm,clip=true]{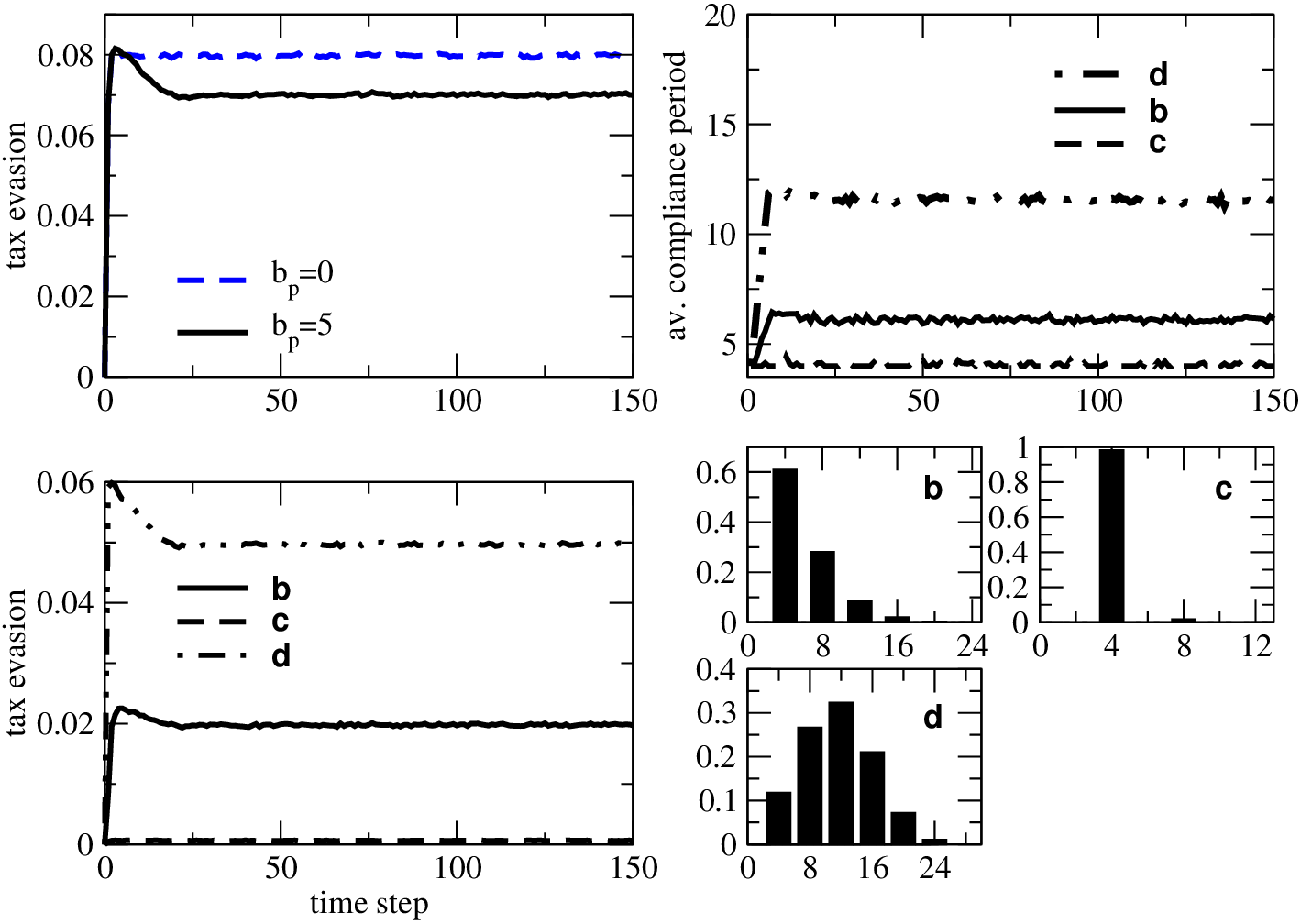}
\end{center}
\caption{Same as Fig. \ref{fig5} but for a society consisting of $35\%$ b-type,
$50\%$ c-type, and $15\%$ d-type agents.}
\label{fig6}
\end{figure}

\begin{figure}[h!]
\begin{center}
\includegraphics[width=8cm,clip=true]{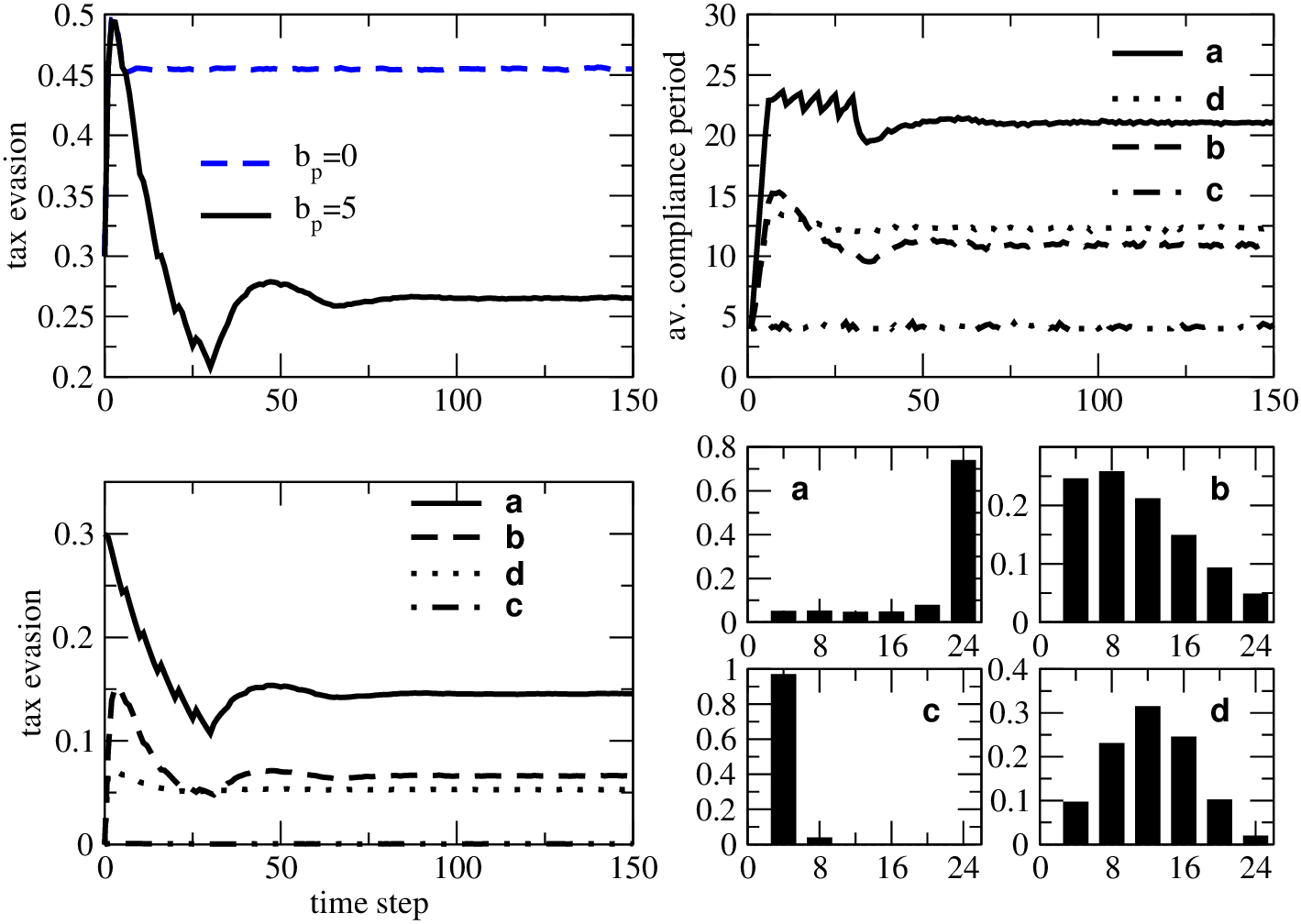}
\end{center}
\caption{Same as Fig. \ref{fig5} but for a society consisting of $30\%$ a-type,
$35\%$ b-type, $20\%$ c-type, and $15\%$ d-type agents.}
\label{fig7}
\end{figure}

The results shown in Figs. \ref{fig5}, \ref{fig6}, \ref{fig7} have been
obtained for societies with a fixed share of $35\%$ b-type and $15\%$ d-type agents.
We then investigate the influence of different population shares of selfish
a-type and ethical c-type agents on the extend of tax evasion and on tax evasion dynamics.

\begin{table}[h!]
\caption{Parameter set for the results shown in Figs. \ref{fig5} to \ref{fig7}.
Note that $T$ and $B$                                                          
are measured in units of $J\equiv 1$.}                                         
\label{tab1}
\begin{center}
\begin{tabular}{|c|r|r|r|r|} \hline
$Agent-type$ & $T_{min}$ & $T_{max}$ & $B_{min}$ & $B_{max}$ \\  \hline\hline
a & 5 & 5 & -20 & -10 \\ \hline
b & 1 & 3 & 0 & 0 \\ \hline
c & 5 & 5 & 10 & 20 \\ \hline
d & 10 & 30 & 0 & 0 \\ \hline                                
\end{tabular}                                                                    
\end{center}                                                                     
\end{table} 

In Fig. \ref{fig5} the percentage of a-type agents is set to $50\%$
and the upper left panel compares the extend of tax evasion and the 
dynamics of tax evasion for backaudit periods $b_p=0$ and $b_p=5$.
The lower left panel reveals the contributions of the individual
agent groups. With respect to the steady state it turns out that 
the a-type agents contribute with $\sim 25\%$ to the overall tax evasion, 
which means that $\sim 50\%$ of a-type agents are evading in agreement with the
result for the homogeneous a-type society in Fig. \ref{fig1}.
Similarly, the d-type agents contribute with $\sim 5\%$ to tax
evasion so that approximately $30\%$ of this group is evading, again
in agreement with the homogeneous result displayed in Fig. \ref{fig4}.
A new feature arises due to the interaction of copying b-type agents
with their neighbors. In fact, we observe a corresponding
contribution of $\sim 10\%$ to overall tax evasion, which means that 
$\sim 28\%$ of b-type agents are evading, compared to $4\%$ in
the homogeneous society (Fig. \ref{fig2}). This is of course due
to the interaction with a-type and to a much smaller extend with
d-type agents. As a result the average compliance period
of b-types reaches now the value of d-types (upper right panel of Fig. 
\ref{fig5}), which, together with the a-types, acquire approximately
the same value than in the homogeneous case.
Also inspection of the forced compliance period (or penalty) distributions (lower right panels
of Fig. \ref{fig5}) reveals a much broader spreading as
compared to Fig. \ref{fig2}, with the maximum at $12$ periods.

The copying feature of b-type agents has much less influence on tax evasion
when we replace the $50\%$ a-type by $50\%$ ethical c-type agents, as
shown in Fig. \ref{fig6}. In fact, because in the stationary state the large
majority of b-types tends to be compliant, the result of Fig. \ref{fig6}
corresponds almost to the result one would obtain by just averaging
the results of Figs. \ref{fig2},\ref{fig3},\ref{fig4}, 
taking into account the parameter distributions.

\begin{figure}[h!]
\begin{center}
\includegraphics[width=8cm,clip=true]{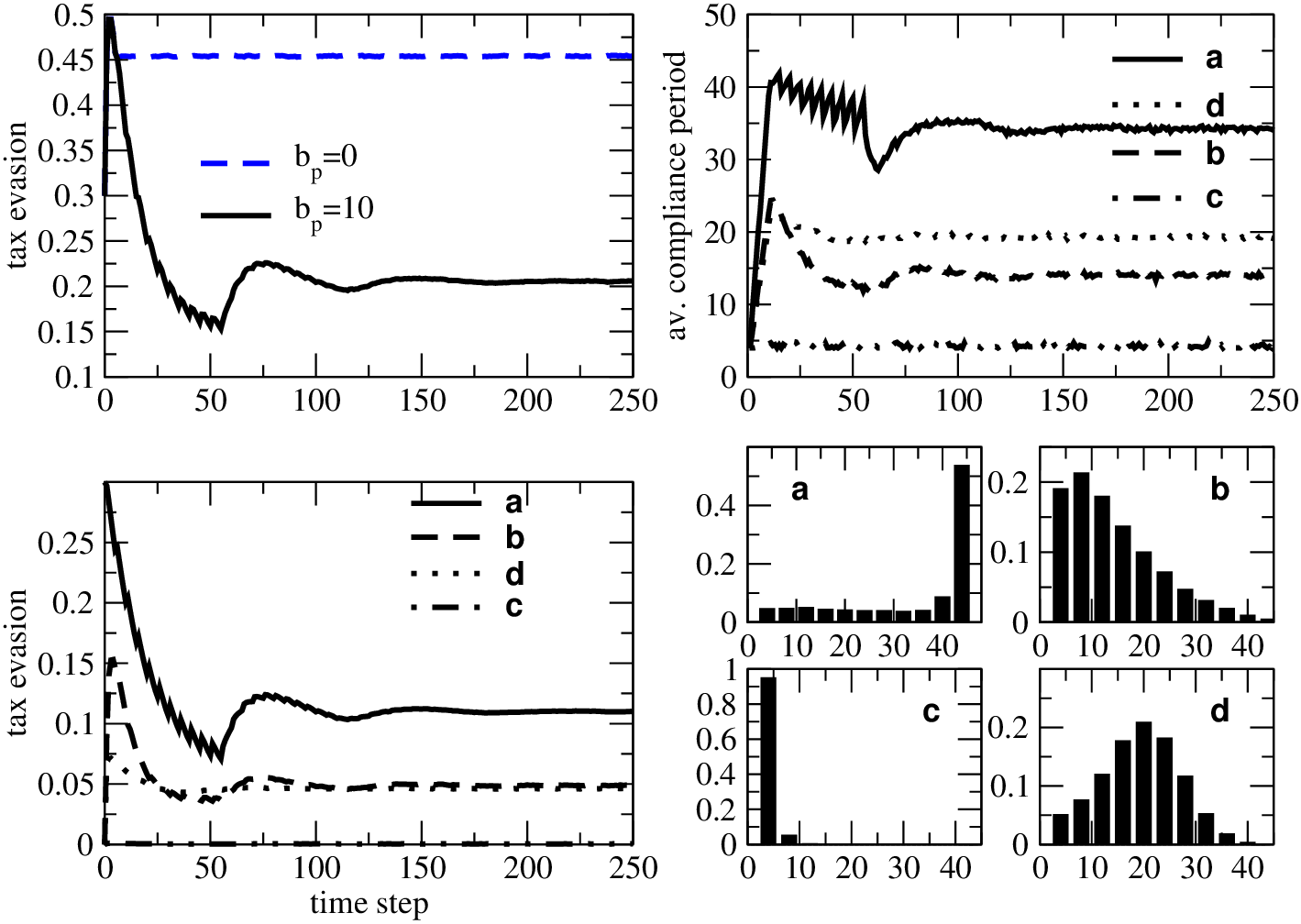}
\end{center}
\caption{Same as Fig. \ref{fig7} but for backaudit period $b_p=10$.}
\label{fig9}
\end{figure}

The case of a society with all four types of 
agents is shown in Fig. \ref{fig7}. Here, those b-types which are the nearest neighbors of a-type agents are
predominantly evading, whereas b-types which are the nearest neighbors
of c-types copy the corresponding compliant behavior. In addition, 
also the fluctuating behavior of d-types is copied by neighboring
b-types, similarly to the previous
figures \ref{fig5}, \ref{fig6}, \ref{fig7}.
As a result, we find that almost $\sim 20\%$ of b-types are evading, a
value slightly lower than in Fig. \ref{fig5}, but still larger than in
the homogeneous case. Inspection of the distribution of
the compliance period for b-type agents yields a compromise between
the situations shown in Fig. \ref{fig5} and Fig. \ref{fig6}. 
In any case, the introduction of backauditing $b_p=5$ reduces the overall
tax evasion from $\sim 45 \%$ to $\sim 27\%$.

Finally, we want to briefly address the influence of the penalty
parameter $h$ as compared to the backaudit period $b_p$. In fact, one
could argue that for a given society, tax evasion for e.g. $h=4$ and $b_p=5$
should be equivalent to $h=2$ and $b_p=11$ since in both cases the
maximum compliance period of detected evadors is 
$k_{max}=(b_p+1)\cdot h=24$. This would be true for societies
of ideal a- or c-type agents which are either always evading
or compliant with probability 'one' (i.e. where $B_i \to \pm \infty$).
However, in our model we consider the case of finite fields and,
moreover, also incorporate b- and d-types with a pronounced distribution
of penalty periods (cf. Figs. \ref{fig2}, \ref{fig4}). For $h=2$ and $h=4$
this  distribution spans the values $k=2,4,6 \dots 24$ and $k=4,8,12 \dots 24$,
respectively. Therefore a given penalty period $k_4$ of the $h=4$ 
computation encompasses two penalty periods ($k_2=k_4$ and $k_2-1$) of the
$h=2$ result and, therefore, shifts the average compliance period
to slightly larger values. As a result, tax evasion for parameters
$h$ and $b_p=4n+3$ is always slighlty smaller than for $2h$ and $b_p=2n+1$
with $n$ integer.

\begin{figure}[h!]
\includegraphics[width=7.5cm,clip=true]{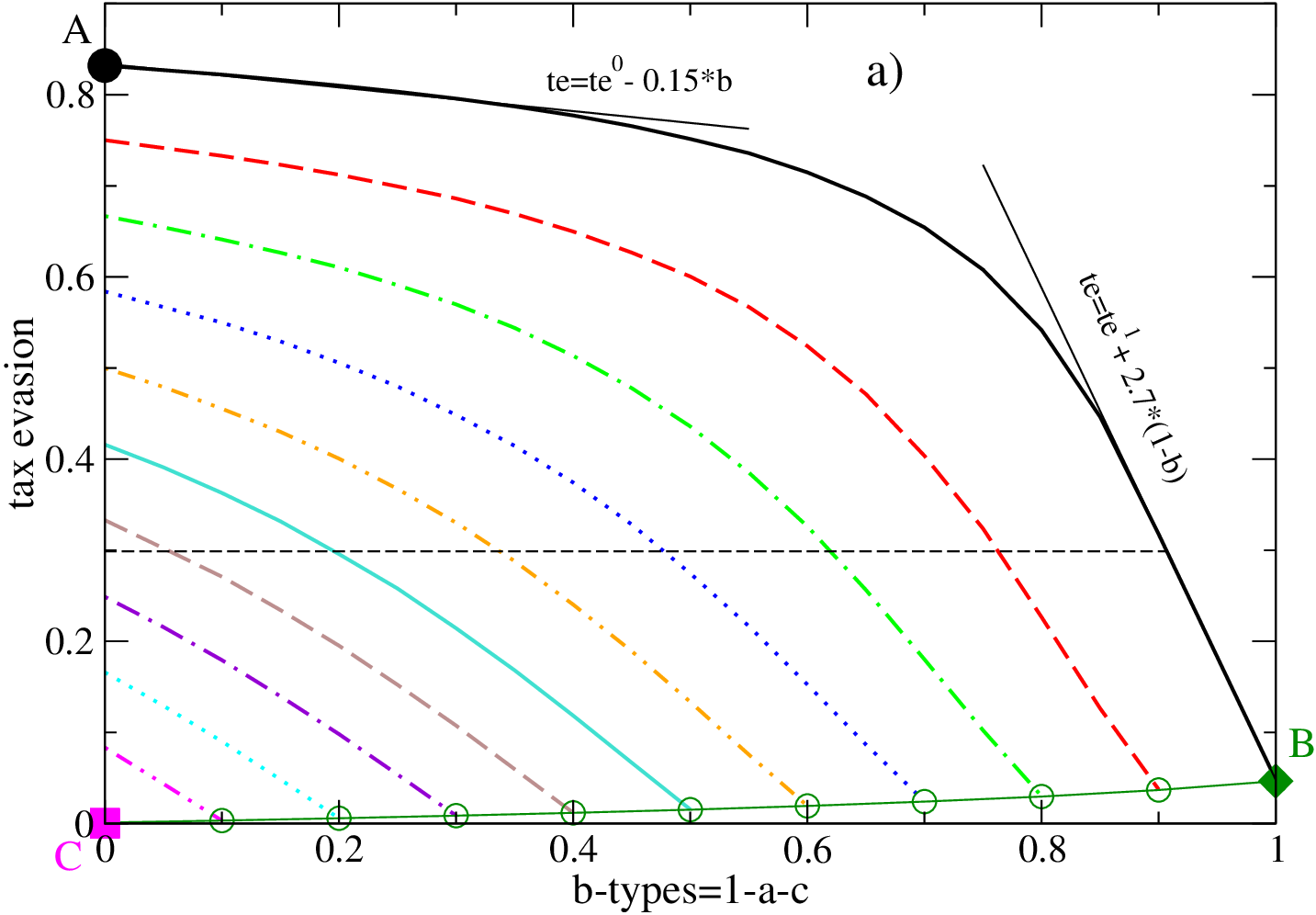}
\includegraphics[width=7.5cm,clip=true]{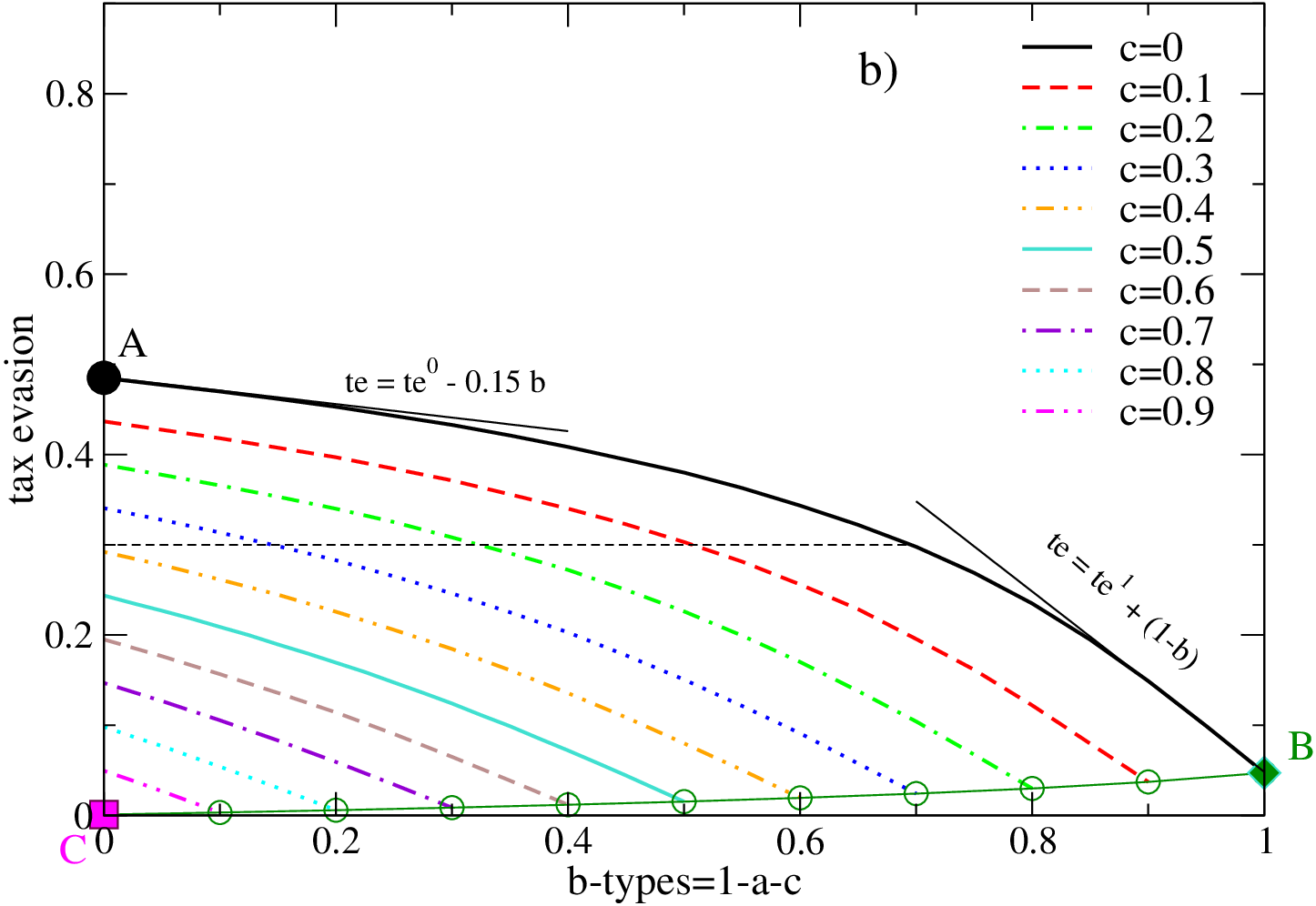}
\includegraphics[width=7.5cm,clip=true]{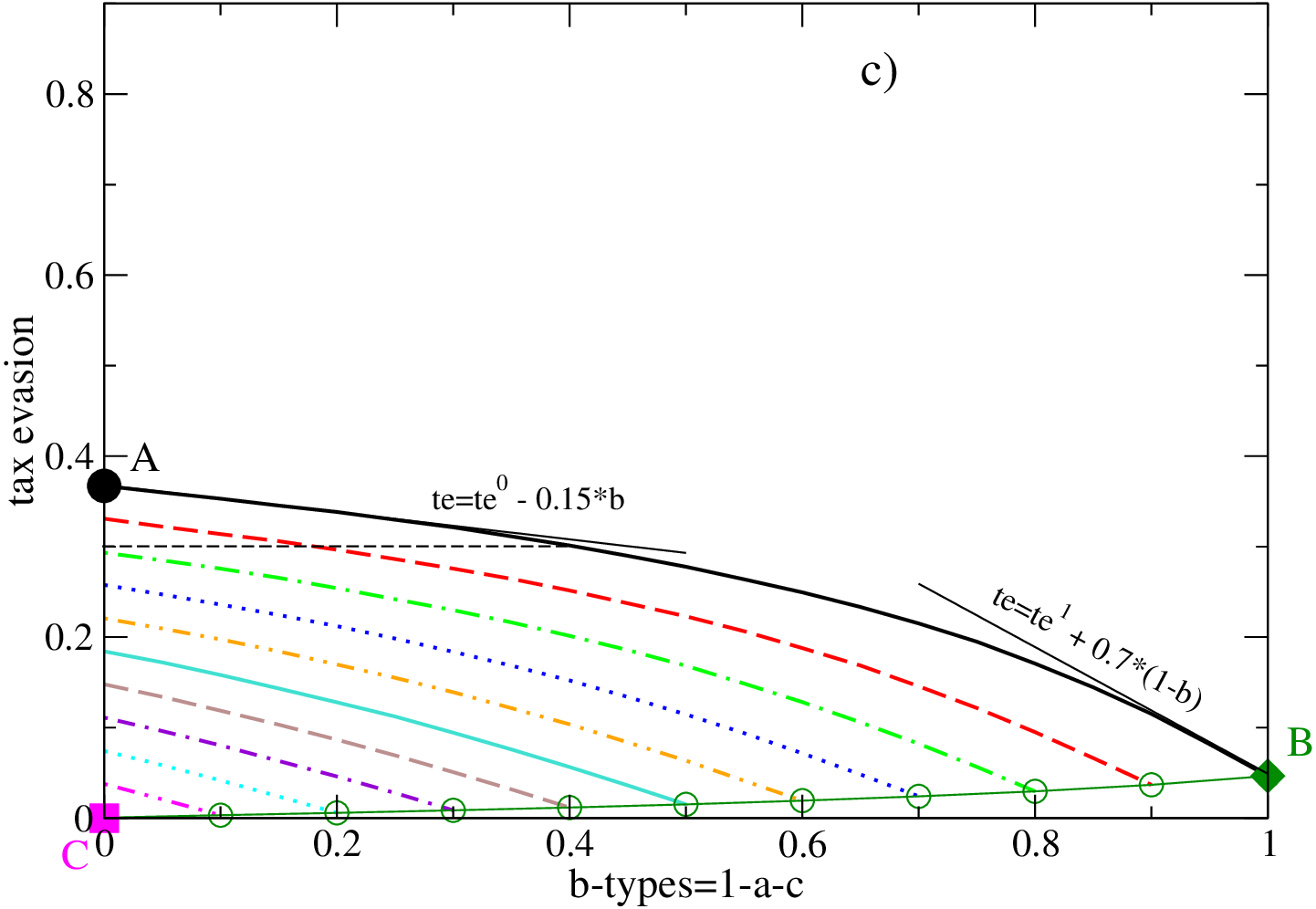}
\caption{Stationary values for the extend of tax evasion 
for audit probability $p_a=5\%$ as a function of 
b-types. The individual curves are for different percentages of
c-type agents and the fraction of d-types is set to zero.
The stationary value is obtained as an average over $30$
time steps after the system is thermalized within $270$ time steps.
We have indicated the tax evasion
for $100\%$ a-types (b-types, c-types) by a solid circle denoted A (diamond denoted B, square denoted C). The thin solid lines tangent to the $c=0$ curve at 
$b=0,1$, respectively, 
visualize the crossover behavior discussed in the text.  
Panel a): No backauditing $b_p=0$, $h=4$.
Panel b): Backauditing $b_p=5$ and $h=4$.   
Panel c): Backauditing $b_p=10$ and $h=4$.
The horizontal dashed line at $30\%$ tax evasion is also discussed in the text.
}
\label{fig8}
\end{figure}

\subsection{Phase diagrams}
In the previous section we have seen that 
copying b-type agents cause the extend of tax evasion of 
heterogeneous societies to be different from the corresponding average
of homogeneous societies. As noted, this is due to the interaction of b-type 
agents with
behaviorally different types of agents in their neighborhood and the actual 
interaction dynamics is
governed by various parameters, including the agent-type distribution in the 
population. 

Therefore, in Figs. \ref{fig8}a,b,c we show the stationary tax evasion 
as a function of the percentage of b-type agents for no backauditing ($b_p=0$)
and backauditing ($b_p=5,10)$, respectively. 
As we are predominantly interested in the antagonistic influence
of a- and c-type agents, the percentage of d-types has been set to zero.
The individual curves in Figs. \ref{fig8}a,b,c represent different percentages
of c-type agents, so that increasing the value on the horizontal axis $b=1-a-c$ 
corresponds to a decrease of the share of a-type agents.

Consider first the $c=0$ curve in Figs. \ref{fig8}a,b,c which displays 
a crossover
behavior (visualized through the thin solid lines) as a function
of b-types. In the vicinity of $b=0$ the vast majority of agents is a-type
and the speckles of b-types tend to copy the a-type behavior. The reason
why tax evasion slightly decreases with the number of b-types is due
to the fact that the latter have a larger probability of staying compliant,
even in an evading environment (cf. table III in Ref. \cite{pick12}). 
On average this probability is $\approx 15\%$ which
coincides with the slope of the $c=0$ curve in Fig. \ref{fig8}
independent of the value of $b_p$.

In contrast, the behavior around $b=1$ can be understood from a society
of predominantly b-type agents with speckles of a-types. In this limit
every a-type contributes to tax evasion proportionally to the
corresponding value of the homogeneous society $te_{hom}(a)$ 
(cf. Fig. \ref{fig1}).
In addition, since each a-type agent is surrounded by four b-types,
the latter have a larger probability of tax evasion and
contribute with $\sim 4a\cdot te_{neighbor}(b)$. The residual fraction
of-types $\sim b-4a$ contributes again with the value of the
homogeneous society $te_{hom}(b)$ (cf. Fig. \ref{fig2}). Note that
due to the coupling to the a-types one has $te_{neighbor}(b) \gg te_{hom}(b)$
and it is exactly due to these neighbors that the slope in Fig. \ref{fig8}a
exceeds the value of 'one'.
Since both $te_{hom}(a)$ and $te_{hom}(b)$ increase with the backaudit
period $b_p$ we find a decrease of slope for the tangent around
$b=1$ going from Fig. \ref{fig8}a to Fig. \ref{fig8}c.

Finally, consider the horizontal dashed line at $30\%$ tax evasion in 
Figs. \ref{fig8}a,b,c. This line shows all combinations of different 
behavioral agent types that are compatible with $30\%$ tax evasion, 
subject to the underlying 
parameter specifications. Inspection of Fig. \ref{fig8} shows that
the increase of the backauditing period not only reduces tax evasion
for a given agent distribution, but also the possible combinations of
agent-types that allow for a specific level of tax evasion (e.g. $30\%$).  
Given that these specifications are implemented 
in a lab experiment with human subjects, it should be possible to derive 
some information on how closely the agent-based model can predict real
human behavior patterns.

\section{Conclusion}\label{sec4}
In this paper we have implemented alternative backauditing schemes, 
in combination with an endogenously penalty, into a multi-agent econophysics model. 
Our investigations essentially confirm the results of a few previous 
analyses on backauditing obtained from 
a lab experiment with human subjects \cite{alm93} and from agent-based models   
of the economics domain \cite{antu07,hopi10}. In particular, our results show 
that the compliance rate may increase substantially with the number 
of backauditing periods, 
subject to the underlying agent type distribution. To this extent, our 
findings support the view that introducing long backauditing periods 
(or lapse of time periods) via tax laws
will effectively reduce the level of tax evasion. 

Naturally, a-type agents due to their inherent selfish behavior are most 
affected by backaudits. In addition, we have shown that through backauditing 
a-types strongly influence the  
endogeneous compliance of b-types, which can be deduced from the b-type
compliance distribution in Figs. \ref{fig5}, \ref{fig6}. 
However, the higher the share of c-types, i.e. the higher the tax morale 
in a society, the less effective a backauditing scheme will be. 
This notwithstanding, 
a given level of tax evasion may be compatible with different
agent type compositions. For example, according
to Fig. \ref{fig8} (dashed horizontal line), a society
with observed $30\%$ tax evasion could contain a share of ethical c-types
ranging from $c=0\%$ up to $c \approx$ $62\%$ (panel a),
$c \approx$ $40\%$ (panel b), and $c \approx$ $20\%$ (panel c), 
subject to the underlying parameter setting.

Finally, the incorporation of behaviorally different agent-types, 
backauditing, and endogenous penalties, adds to the microfoundation of 
econophysics models and allows for   
comparing results obtained from such a model with those obtained 
from human-subject-based lab experiments. As noted, this may permit an
empirical determination of the parameters of our model and, 
in turn, a prediction of the extend of tax evasion under 
different parameter values. But, of course, this deliniates a future reserach agenda.


\begin{thebibliography}{00}
\bibitem{note1} This differs from the case where a  temperature gradient
is applied to the system boundaries and local $T_i$'s are then obtained as a 
result of heat flow as e.g. in Refs. \cite{harris88,agli10}.
\bibitem{note2} With an initial condition where all a-types are set to 
                compliance ($S_i=+1$) tax evasion would start at $0\%$ 
                at time step $'0'$ but from the first time step continue
                to follow the same curve shown in Fig. \ref{fig1} for
                full non-compliance as initial condition.
\end{thebibliography}
\end{document}